\documentclass[prd,twocolumn,a4paper,preprintnumbers,showpacs,byrevtex3,final,nofootinbib]{revtex4}

\usepackage{epsfig, amssymb, amsmath}
\DeclareGraphicsExtensions{.jpg, .eps}
\newcommand{\be}{\begin{equation}}
\newcommand{\ee}{\end{equation}}
\newcommand{\bea}{\begin{eqnarray}}
\newcommand{\eea}{\end{eqnarray}}
\newcommand{\ba}{\begin{array}}
\newcommand{\ea}{\end{array}}

\def\bbox{{\,\lower0.9pt\vbox{\hrule \hbox{\vrule height 0.2 cm
\hskip 0.2 cm \vrule height 0.2 cm}\hrule}\,}}
\newcommand{\dsl}{\pa \kern-0.5em /}

\newcommand{\nn}{\nonumber \\}

\font\mybb=msbm10 at 10pt
\def\bb#1{\hbox{\mybb#1}}

\def\bR {\bb{R}}

\def\dd{\text{d}}

\def\appendix#1{
  \addtocounter{section}{1}
  \setcounter{equation}{0}
  \renewcommand{\thesection}{\Alph{section}}
  \section*{Appendix \thesection\protect\indent \parbox[t]{11.15cm}
  {#1} }
  \addcontentsline{toc}{section}{Appendix \thesection\ \ \ #1}
  }

\begin{document}

\title{Recurrent Acceleration in Axion Dilaton Cosmology}
\author{Julian Sonner}
\author{Paul K. Townsend}
\affiliation{Department of Applied Mathematics and Theoretical Physics \\ Centre for Mathematical Sciences, University of Cambridge \\ Wilberforce Road, Cambridge, CB3 0WA, UK }

\preprint{DAMTP-2006-63}

\begin{abstract}A class of Einstein-dilaton-axion models is found for which 
almost all  flat expanding homogeneous and isotropic universes 
undergo recurrent periods of acceleration.  We also extend recent 
results on eternally accelerating 
open universes.
\end{abstract}


\vskip 1cm
\vskip 1cm















\maketitle


\setcounter{page}{1}
\section{Introduction} \setcounter{equation}{0}

The discovery that the expansion of our Universe is accelerating, 
presumably due to ``dark energy'', together with the compatibility 
of recent astronomical observations 
with the hypothesis of an earlier inflationary epoch, has given a new 
impetus to studies of cosmologies driven by scalar fields. Models 
with a single scalar field have a long history. In more recent times, 
models with two or more scalars have attracted much attention;
in particular models with both dilaton and axion \cite{Linde:1991km}.
Of most  relevance here is the work of  Billyard et al. \cite{Billyard:2000cz} 
because the models we study  include one  studied by them, and our methods 
are similar.  

We suppose that a dilaton field $\sigma$ and axion field 
$\chi$, which together define a 
map from a $D$-dimensional spacetime to a hyperbolic space with 
an $Sl(2;\bR)$-invariant metric, are coupled to gravity. The 
$Sl(2;\bR)$-invariance must be broken by any non-constant scalar 
field potential, but if we require only invariance under dilatations 
then an exponential potential for the dilaton is allowed. These 
considerations lead to a low-energy Lagrangian density of the form 
\be\label{laginitial}
{\cal L} = \sqrt{-\det g}\left[ R -
{1\over2}\left(\partial\sigma\right)^2 - 
{1\over2} e^{\mu\sigma}\left(\partial\chi\right)^2 - 
\Lambda e^{-\lambda\sigma} \right]
\ee
where $\mu$ and $\lambda$ are constants, as is the `cosmological 
constant' $\Lambda$, which we here assume to be positive. We may 
choose $\lambda\ge0$ without loss of generality but then the sign 
of $\mu$ is significant, so we allow for either sign. Note that 
$|\mu|$ is proportional to the radius of the hyperbolic target 
space. When $\mu=0$, the target space is flat and the axion field 
decouples; the resulting cosmologies were studied 
in \cite{Bergshoeff:2003vb,Townsend:2004zp,Collinucci:2004iw} so we assume here 
that $\mu\ne0$. Models of this type can be motivated in various ways. 
For example the Freedman-Schwarz $D=4$ supergravity 
theory \cite{Freedman:1978ra} has a consistent truncation 
to (\ref{laginitial}) with $2\lambda=-\mu =2$. 

We shall study flat FLRW cosmological solutions of the above model  
as a function of the parameters $\mu$ and $\lambda$. The $D=4$ model 
with $\lambda=-\mu=2$ was previously studied by Billyard et al. 
\cite{Billyard:2000cz} with a string-theory motivation that led 
them to consider a `string-frame' metric rather than the 
Einstein-frame metric implicit in (\ref{laginitial}). They 
found an interesting quasi-cyclic evolution of flat cosmological 
models that they interpreted as successive periods of expansion 
and contraction. As we point out here, the Einstein-frame universe 
is either eternally expanding or eternally contracting, but each 
undergoes successive periods of {\it acceleration}. The  
`string-frame'  cosmologies of 
\cite{Billyard:2000cz} also undergo successive periods of acceleration 
because  this is an immediate consequence of  re-expansion following a 
period of  contraction, but acceleration in non-Einstein frames does 
not  imply acceleration in the Einstein frame; in particular, a 
flat Einstein-frame universe cannot smoothly pass from  contraction 
to expansion (see e.g. \cite{Khoury:2001bz}) so this source of 
cosmic acceleration is not available in the Einstein frame. 
We shall show that, in Einstein frame, almost all flat  
expanding FLRW cosmologies undergo quasi-cyclic alternation 
between periods of  acceleration and deceleration provided that
both $\lambda$ and $-\mu$ are  sufficiently large. The late-time behaviour 
is a power-law expansion which may be accelerating or decelerating, depending on the value of 
$\mu/\lambda$.

Of  course, there is no physical significance to the choice of frame 
within the models we consider  but this choice  becomes significant 
once one considers the additional fields needed for a realistic 
theory.  In this larger context, the choice of Einstein frame is equivalent to the assumption that  the
energy-momentum-stress tensor of the matter fields responsible for structure
couple to the Einstein-frame metric, in the sense that the motion of a free
particle associated with such a field is geodesic in this frame, an
assumption that  is consistent with observations.

The model defined by (\ref{laginitial}) can be consistently truncated 
by taking $\chi=\chi_0$, for any constant $\chi_0$.  Solutions of the 
resulting Einstein-dilaton  model are therefore also solutions of the 
Einstein-dilaton-axion model, and this  provides a useful check  
on our results. The Einstein-dilaton model has been much studied 
in the past, but some features in relation to eternal cosmic 
acceleration of open FLRW universes in models arising from hyperbolic 
compactification have come to light only recently. For the values of $\lambda$ 
that arise in such compactifications,  Chen et al. \cite{Chen:2003dc} have shown 
that  late-time `eternal'  acceleration is possible for open universes
(see also \cite{Neupane:2003cs}).  Andersson and Heinzle have further
shown that `eternal' acceleration is possible in the stronger sense  of  ``at 
all times''  (and not merely in the future of some 
initial onset of  acceleration), and that there is a unique open universe
cosmology with this property \cite{Andersson:2006du}. Here we extend this result, 
in the  form of a conjecture, to other values of $\lambda$, including those that arise
from flux compactifications.

\section{Preliminaries}
 \setcounter{equation}{0}
We consider a $D$-dimensional  FLRW spacetime with  metric
\be\label{eq:ansatz}
\dd s^2_D = - e^{2\alpha\varphi(\tau)} f^2(\tau) \dd\tau^2 + 
e^{2\beta\varphi(\tau)} \dd\Sigma_k^2
\ee
where $\dd\Sigma_k^2$  is the metric for a $(D-1)$-dimensional 
maximally symmetric space of  constant curvature $k=-1,0,1$, and $\alpha,\beta$ are the $D$-dependent constants
\be
\alpha = \sqrt{(D-1)\over 2(D-2)}\, , \qquad \beta = 1/\sqrt{2(D-1)(D-2)}\, . 
\ee
Supposing that $\sigma$ and $\chi$ are functions only of $\tau$, we find that the field equations reduce to equations of motion, and a constraint, for the variables $(\varphi,\sigma,\chi)$ that are the Euler-Lagrange equations of the effective Lagrangian
\begin{multline}
L_{eff}= {1\over2} \left[ f^{-1} \left(\dot\varphi^2 -  \dot\sigma^2 - e^{\mu\sigma} \dot \chi^2 \right)\right.\\\left. - fe^{2\alpha\varphi}\left(k\beta^{-2}e^{-2\beta\varphi}-2\Lambda e^{-\lambda \sigma}\right)\right]\, 
\end{multline}
where the overdot indicates differentiation with respect to $\tau$. We may take $f$ to be any positive function. We may also make any choice of the positive constant $\Lambda$ since its value is changed  by a shift  of $\sigma$. It is convenient to choose
\be\label{gc}
f= e^{-\alpha\varphi + {1\over2}\lambda\sigma}\, , \qquad 2\Lambda=1\, , 
\ee
and to define 
\be
u=\dot\sigma\, ,\qquad v= \dot\varphi\, . 
\ee
In this case, the equations of motion following from the effective
lagrangian are equivalent to the coupled  ordinary differential equations
\begin{align}
\dot u &= -\alpha u v + \frac{1}{2}(\lambda - \mu)(u^2+1) + 
\frac{1}{2}\mu v^2 + \frac{\mu k}{2\beta^2}\, e^{\lambda\sigma -2\beta\varphi}\nonumber \\
\dot v &= -\alpha v^2 + \frac{1}{2}\lambda uv +\alpha -
\frac{k}{2\alpha\beta^2}\, e^{\lambda\sigma - 2 \beta\varphi}\, , 
\label{ODEs}
\end{align}
together with the constraint
\be\label{gencon}
\dot \chi^2 = e^{-\mu\sigma}\left[ v^2 - u^2 - 1
  + \frac{k}{\beta^2}e^{\lambda\sigma - 2 \beta\varphi}\right]\,.
\ee
The $\chi$ equation of motion is implied by differentiating the constraint.

The spacetime metric can be put in the standard FLRW form
\be\label{FLRW}
\dd s^2 =-\dd t^2 + S^2(t)\, \dd\Sigma_k^2
\ee
by defining
\be
\dd t = e^{{1\over 2}\lambda\sigma}\dd\tau\, ,\qquad S(t) = e^{\beta\varphi(\tau)}\, . 
\ee
The condition for the universe to be expanding is $dS/dt>0$, which is equivalent to $\dot \varphi>0$.
The condition for positive acceleration is $d^2S/dt^2>0$, which is equivalent to 
\be\label{eq:accel_condition}
v^2 < 2\alpha^2 - \frac{k}{\beta^2} e^{\lambda\sigma -2\beta\varphi}\, . 
\ee
Using the constraint to eliminate the $k$-dependent term, one finds that the acceleration condition
is equivalent to the inequality
\be\label{eq:accel_condition2}
\frac{1}{2}\left[\dot\sigma^2 + e^{\mu\sigma}\dot\chi^2\right] < \alpha\beta\, .  
\ee
This has a simple physical interpretation because the left hand side is the kinetic energy of the 
scalar fields. 

We will now consider in turn the two special cases of (i) constant axion and (ii) $k=0$. In these two 
cases the problem can be reduced to the analysis of a 2-dimensional autonomous dynamical system. 
For both cases, it is convenient to define 
\be\label{lambdas}
\lambda_c = 2\sqrt{\alpha\beta} = \sqrt{\frac{2}{D-2}}\, ,\qquad
\lambda_h = 2\alpha = \sqrt{D-1}\, \lambda_c \, . 
\ee
These quantities have a general convention-independent significance, but take the above particular values in our conventions.


\section{Constant Axion} \setcounter{equation}{0}
It is consistent to seek solutions of (\ref{ODEs}) and (\ref{gencon}) by setting $\dot\chi=0$, in which case 
it is convenient to use the constraint to eliminate the $k$-dependent terms from the equations of motion. 
The resulting equations of motion  are 
\bea\label{eq:chiiszerosytem}
\dot u &=& -\alpha uv + \frac{1}{2}\lambda (u^2 +1)\, ,\nonumber\\
\dot v &=& -\beta (v^2 - 1) - \frac{1}{2\alpha}u^2 + \frac{1}{2}\lambda u v\, ,
\eea
and the constraint  is
\be
u^2 - v^2 +1 = \frac{k}{\beta^2}e^{\lambda\sigma - 2 \beta\varphi}\,.
\ee
Equations (\ref{eq:chiiszerosytem}) define a two-dimensional autonomous system, analyzed by Halliwell  for $D=4$ \cite{Halliwell:1986ja}.  The generalization to arbitrary $D$ was given in \cite{Townsend:2004zp}. In these references, the phase portraits were presented in the $(u,v)$ plane.  Here we summarize the  results of a {\it global} phase plane analysis \cite{SonnerTownsend2006I},  paying particular attention to cosmic acceleration; setting $\dot\chi=0$ in (\ref{eq:accel_condition2}), we find that acceleration occurs when
\be
u^2 < 2\alpha\beta\, . 
\ee

By using alternative phase plane coordinates one can study the nature of the trajectories at infinity, fixed points in particular.  Actually, since trajectories at  infinity are the same for positive and negative $\Lambda$, the phase-plane portrait for $\Lambda>0$ can be joined to the phase-plane portrait for $\Lambda<0$ at infinity. The resulting `total' phase plane is topologically a sphere.  The fixed points on the sphere have positions that are functions of $\lambda$ and  they all are hyperbolic for finite (non-negative) $\lambda$ except when 
\begin{itemize}
\item (i) $\lambda= \lambda_c$. The $k=0$ and $k\ne0$ fixed points in the $(u,v)$ plane coincide at this  `critical' value of $\lambda$, which is a transcritical bifurcation point in  the family of dynamical systems parametrized by $\lambda$ \cite{Sonner:2005sj}. 

\item (ii) $\lambda= \lambda_h$. This was called the `hypercritical' point in \cite{Townsend:2004zp}.
The $k=0$ fixed point goes to infinity at this value of $\lambda$. Globally, this value of $\lambda$ is another transcritical bifurcation \cite{SonnerTownsend2006I}.
\end{itemize}

\begin{figure}[h!]
  
      \begin{center}  
      \epsfig{file=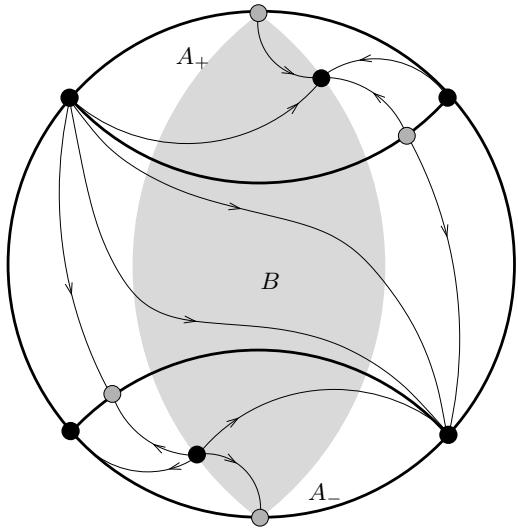, scale=0.35}
      \begin{picture}(0,0)(0.01,0.01)
	\put(-132,175){$A_+$}
	\put(-82,10){$A_-$}
	\put(-100,90){$B$}
	\end{picture}
      \caption{$\lambda_c< \lambda \le \bar\lambda \le \lambda_h$, and
      $D\le9$. The acceleration region is shaded.  Regions $A_\pm$
      correspond to open universes that are either expanding ($A_+$)
      or contracting ($A_-$). Region $B$ corresponds to closed universes. \label{fig:global1}}
      
    \end{center}
  \end{figure}

Here we are concerned only with the hemisphere that contains the  trajectories of the $\Lambda>0$ model, including the circle at infinity. This is illustrated for $\lambda>\lambda_c$ in Figs. 1\&2, where the shaded region is  the acceleration region. Fig. 1 applies for $\lambda<\lambda_h$ and Fig. 2 for $\lambda>\lambda_h$.  These two cases arise naturally in compactifications from general relativity in a higher dimension, with $\lambda<\lambda_h$ arising from compactification on hyperbolic spaces and $\lambda>\lambda_h$ arising from flux compactifications (see e.g. \cite{Townsend:2003qv}). 
Trajectories in  region $A_+$ correspond to expanding open ($k=-1)$ universes, while those in region $A_-$ correspond to contracting open universes. Region $B$ trajectories correspond to closed ($k=1$) universes, which may be contracting or expanding. Note that there are two fixed points  that are simultaneously in $A_+$ and on the boundary of the acceleration region. The fixed point at infinity is a saddle and there is a separatrix trajectory that connects it to the other of these two fixed points, which is either a stable focus or a stable node depending on both $\lambda$ and $D$. This is a general feature for $\lambda>\lambda_c$,  which we assume in the following.   As shown in \cite{Townsend:2004zp}, this $k=-1$ fixed point
 at finite $(u,v)$ is a stable focus when both $D\le9$ {\it and} 
\be
\lambda> \bar\lambda \equiv 
\frac{4}{\sqrt{(10-D)(D-2)}}\, .
\ee 
Otherwise it is a stable node.  Note that 
\be
\lambda_c < \bar\lambda \leq \lambda_h
\ee
where the second inequality is saturated for $D=9$.


 \begin{figure}[h!]
      \begin{center}  
      \epsfig{file=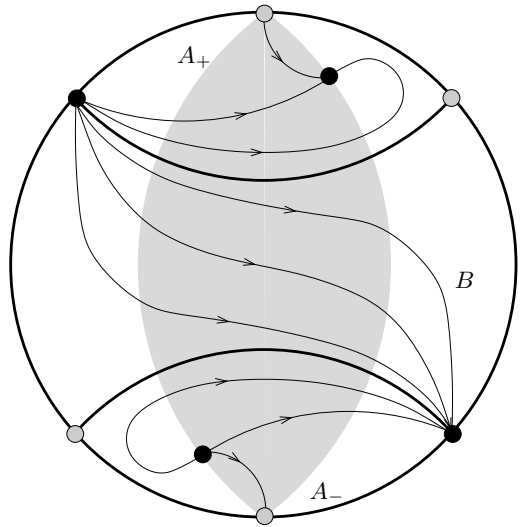, scale=0.35}
       \begin{picture}(0,0)(0.01,0.01)
	\put(-132,175){$A_+$}
	\put(-82,10){$A_-$}
	\put(-27,90){$B$}
	\end{picture}
      \caption{$\lambda >\lambda_h$, and $D\ge10$.  The acceleration region is shaded. Note the (separatrix) trajectory of region $A_+$ that lies entirely within the acceleration region.
      \label{fig:eternal_accel_I}}
      
    \end{center}
  \end{figure}


A node is diffeomorphic to a focus and, consequently, they are often regarded as equivalent in texts on dynamical systems.  However, the distinction is physically relevant in the present context because a diffeomorphism would alter the boundary of the acceleration region. In the approach to a focus on the boundary of this region a trajectory will ultimately cycle in and out of the region. In contrast, in the approach to a node on the boundary of the acceleration region, a generic trajectory will ultimately 
be continuously inside or continuously outside the region. Moreover, for a node the possibility 
arises that the above-mentioned separatrix trajectory will lie entirely in the acceleration region; no
other trajectory can have this property because the fixed point at infinity on the acceleration boundary is a  saddle.  A natural conjecture is that the separatrix trajectory lies entirely inside the acceleration region 
if and only if the `internal' fixed point is a node rather than a focus. This is illustrated for $\lambda<\lambda_h$ in Fig. 1,  and for  $\lambda>\lambda_h$ in Fig. 2. 

This conjecture is consistent with recent results of  Andersson and Heinzle \cite{Andersson:2006du}, as we now explain. These authors considered open universes in $D=1+m$ dimensions ($m\ge2$) arising from compactification on an $n$-dimensional compact hyperbolic space. The equations that they consider are the same as those deduced from an effective $D$-dimensional Einstein-dilaton 
theory with (see e.g. \cite{Bremer:1998zp})
\be
\lambda = \sqrt{\frac{2(m+n-1)}{n(m-1)} }\equiv \lambda_{AH}\, , \qquad (n\ge2).
\ee
Note that $\lambda_c<\lambda_{AH}<\lambda_h$, so the fixed points are as shown in Fig. 1 rather than Fig. 2. The
global phase space of Andersson and Heinzle is  region $A_+$ in Fig. 1. They proved that the separatrix
trajectory lies entirely within the acceleration region if and only if $n+m+1 \ge 10$. This result is consistent with our conjecture (and proves it for some cases). To see this, we first note that the condition
$n+m+1 \ge 10$ is automatically satisfied if $m\ge9$, but then $D\ge10$ and it is also true that the  `internal' fixed point is a node. For $m\le8$, we have a node only if $\lambda<\bar\lambda$, but
one can show that
\be
\lambda_{AH}\leq \bar\lambda  \ \Leftrightarrow \  \frac{2\left(9-n-m\right)}{n(9-m)} \leq 0 
\ee
and hence that the `internal' fixed point is a node if and only if $n+m+1 \ge 10$. Thus, our conjecture is supported by the results of \cite{Andersson:2006du} but predicts the existence of an eternally accelerating trajectory for $D\le8$ as long as $\lambda>\bar\lambda$, and also for $\lambda>\lambda_h$ when $D\ge10$, as shown in Fig. 2.  As mentioned above, this latter case is applicable to flux compactifications.

It should be noted that  eternal acceleration, in the above sense, contradicts observations, which also 
favour a flat universe, so eternally accelerating $k=-1$ universes are of little astronomical  interest. For the remainder of the paper we consider only flat universes, and we seek solutions 
that undergo multiple periods of acceleration.


\section{Varying axion: flat cosmologies} \setcounter{equation}{0}
Another case in which the dynamics is reducible to that of a
two-dimensional autonomous dynamical system is that of spatially flat 
cosmologies, but now allowing for arbitrary motion on the dilaton-axion 
target space. Setting $k=0$ in (\ref{ODEs}) we have the equations
\bea\label{DS2}
\dot u &=& -\alpha uv +{1\over2}\left(\lambda-\mu\right)(u^2+1) + 
{1\over2}\mu v^2 \, , \nonumber\\
\dot v &=& -\alpha v^2 + {1\over2}\lambda uv +\alpha\, , 
\eea
which define a 2-dimensional autonomous dynamical system. Setting $k=0$ in 
(\ref{gencon}) we have the constraint 
\be\label{constraint}
\dot\chi^2 = e^{-\mu\sigma}\left[ v^2-u^2 - 1\right]\, , 
\ee
which restricts the physical  phase space by the inequality
\be\label{ineq}
v^2-u^2 - 1\ge0\, . 
\ee
Note that this is a non-compact region of the $(u,v)$ plane. Before proceeding,  it is convenient to introduce new variables that will allow a compactification of the phase space and thus facilitate the analysis of any fixed points at infinity. We choose these `global' variables to be\footnote{The phase portrait is invariant (up to direction of the trajectories) under reflection about the 
$x$-axis so the variables $(x,\tilde y)$ with $\tilde y= y^2$ constitute another  possible choice of global variables; this choice was made in \cite{Billyard:2000cz} but our choice makes the issues
of expansion and acceleration clearer.}
\be
x= u v^{-1}\, ,\quad y= v^{-1}\qquad\qquad u =
xy^{-1}\,,\quad v = y^{-1}\,.
\ee
The constraint now implies that $x^2 +y^2 \le1 $, so that the allowed region of the $(u,v)$ phase plane has been mapped to the unit disk in the $(x,y)$ plane.  If we define a new independent  variable $\hat \tau$ such that
\be
|v| d \tau = d\hat \tau\,, 
\ee
then the  equations that define the dynamical system in the $(x,y)$ variables are
\bea
\frac{d x}{d \hat \tau} &=&   \text{sgn}(y)\left\{y^2\left[\frac{1}{2}(\lambda-\mu)     -\alpha x \right] +\frac{\mu}{2}(1-x^2)\right\}\nn
\frac{d y}{d \hat \tau} &=&  |y|\left\{\alpha -{1\over2}\lambda x -\alpha y^2\right\}\, . 
\eea
The sign of $y$ is fixed for any given trajectory, so these equations define {\it two} independent dynamical systems, one for $y>0$ and another for $y<0$. As the condition for expansion in the new variables is $y>0$, trajectories with $y>0$ correspond to expanding universes while those with $y<0$ correspond to contracting universes, and time reversal takes one set of trajectories into the other\footnote{Each of these two dynamical systems can be compactified by the addition of a $y=0$ trajectory but the two $y=0$ trajectories cannot be identified because their directions are opposed. The full compactified phase space is therefore two disjoint hemicircles, as shown in Figures 3 and 4.}. The condition for  positive acceleration (which is unchanged by time reversal) is (\ref{eq:accel_condition}), but the equation of motion for $\dot v$ can be used to simplify this to $v^2 <2\alpha^2$, which is equivalent to
\be
y^2 >2/ \lambda_h^2 \, . 
\ee

Fixed points associated with infinite values 
of $u$ and $v$ have the following $(x,y)$ coordinates, and eigenvalues:
\be
\left(\pm 1,0  \right) : \left\{\alpha \mp \frac{\lambda}{2},\mp \mu
\right\}\, .
\ee
The nature of the fixed points at infinity therefore depends on the values of $\lambda$ and $\mu$.
Excluding $\mu=0$ and $\lambda\ne\lambda_h$, which yield non-hyperbolic fixed points, 
we have:
\begin{itemize}
\item $\lambda < \lambda_h$, $\mu>0$: $(1,0)$ is a saddle,  and  
$(-1,0)$ is an unstable node. 
\item $\lambda <\lambda_h$, $\mu<0$: $(1,0)$ is an 
unstable node, and $(-1,0)$ is a  saddle.

\item $\lambda>\lambda_h$, $\mu>0$: $(1,0)$ is a stable node, 
and $(-1,0)$ is a  saddle.

\item $\lambda>\lambda_h$, $\mu<0$: both $(1,0)$ and  
$(-1,0)$ are saddles. 
\end{itemize}

For $\lambda < \lambda_h$ there is a pair of fixed points, not at infinity, on the boundary of the allowed region, and hence on the unit circle in the $(x,y)$-plane; one has $y>0$ and the other $y<0$. 
The $(x,y)$ coordinates, and eigenvalues, of the $y>0$ fixed point, are 
\be
\left(\frac{\lambda}{\lambda_h}\, , \, \sqrt{1-\frac{\lambda^2}{\lambda_h^2}}  \right)
\ : \ \left\{\frac{\lambda^2-\lambda_h^2}{2\lambda_h}\, , \, 
\frac{\lambda(\mu_c-\mu) }{\lambda_h}  \right\}\,,
\ee
respectively, where
\be
\mu_c  = \frac{1}{\lambda} \left(\lambda^2 - \lambda_h^2 \right)\,.
\ee
This type of fixed point is a stable node for $\mu>\mu_c$ and a 
saddle for $\mu<\mu_c$. It is non-hyperbolic on the curve  $\mu=\mu_c(\lambda)$, 
but we will leave consideration of these cases to another 
article \cite{SonnerTownsend2006I}.  We note, however, that the
Freedman-Schwarz supergravity model mentioned in the introduction has
 $\mu=\mu_c$, with $\lambda=\lambda_c$,  so this is an example of a model with 
non-hyperbolic fixed points. 

Finally, if $\mu(\mu-\lambda)>0$ then there is a fixed point for $y>0$ in the interior of the unit circle, and another for $y<0$. The  $(x,y)$ coordinates,  and eigenvalues, of the $y>0$ fixed point are
\be
\left(\frac{\lambda_h}{\lambda-\mu}\, , \,   \sqrt{\frac{\mu}{\mu-\lambda}}
\right) \ : \ \left\{E_+, E_-\right\}\, , 
\ee
respectively, where
\be\label{evaluespm}
E_\pm = -\frac{\mu}{2(\mu - \lambda)}\left( \alpha \pm 
\sqrt{\Delta/\mu}\right)
\ee
with
\be\label{eq:defdelta}
\Delta(\lambda,\mu) =\alpha^2 \mu_c + 2\lambda (\mu-\mu_c)^2 - 7
\alpha^2 (\mu-\mu_c)\,.
\ee
Given that this `interior' fixed point exists, which it will do only if $\mu(\mu-\lambda)>0$, then it lies within the allowed region only if $\mu<\mu_c$. Together, these two conditions are equivalent to 
\begin{equation}\label{eq:condinf2}
\mu < \text{inf}\,[0,\mu_c]\,.
\end{equation}
As this implies that $\mu<0$, the fixed point will be a stable focus whenever $\Delta>0$.

At the `interior' fixed point we may  integrate the equations $\dot\sigma = u$ and $\dot\varphi = v$ 
to get the solution
\begin{align}
\sigma &= \frac{2\alpha}{\sqrt{\mu (\mu - \lambda)}}(\tau -
\tau_0)\,, \qquad\varphi = \sqrt{\frac{\mu - \lambda}{\mu}} (\tau - \tau_0) + \varphi_0\,,\nonumber\\
\chi &= \pm\frac{ \sqrt{ \lambda(\mu_c - \mu) } } {\alpha\mu}\,e^{-\frac{\alpha\mu}{\sqrt{\mu(\mu-\lambda)}}\,(\tau - \tau_0)} + \chi_0\,.
\end{align}
By an appropriate choice of $\varphi_0$, the metric can be 
brought  to the standard FLRW form
(\ref{FLRW} ) with scale factor
\be\label{scalefix}
S = t^{\delta}\, , \qquad \delta = \frac{\lambda - \mu}{\lambda(D-1)}\,, 
\ee
where the FLRW time is
\be
t\propto  \exp\left[ \frac{\lambda
    \alpha(\tau - \tau_0)}{\sqrt{\mu(\mu- \lambda)}}\, \right]\, . 
\ee
This fixed point lies inside the acceleration region if $\delta>1$, which is equivalent to
\begin{equation}\label{eq:spiral_condition}
\mu/\lambda < - 2/\lambda_c^2\,.
\end{equation}
Note that $\delta\to\infty$ as $\mu\to -\infty$, so that the acceleration at the fixed point can 
be arbitrarily large.  

The implications of this  `interior'  fixed point solution for the other $k=0$ trajectories is what will mostly concern us for the remainder of this article. The phase portrait depends on whether there is also a fixed point on the unit circle, which there will be only if $\lambda<\lambda_h$.  We will consider in turn the two cases $\lambda>\lambda_h$ and $\lambda<\lambda_h$. 

\subsection{$\lambda>\lambda_h$: Recurrent Acceleration} 
For $\lambda>\lambda_h$, we have $\mu_c>0$, so that (\ref{eq:condinf2}) becomes $\mu<0$.
Since $\mu<\mu_c$, equation (\ref{eq:defdelta}) implies that $\Delta>0$. The `interior'  fixed point is therefore a stable focus. For $\mu<0$ the fixed points at infinity are both saddles, so  the `trajectory'  at infinity, $y=0$, and the `boundary'  trajectory $x^2+y^2=1$, with $y>0$, constitute a heteroclinic cycle enclosing the stable focus. All other $k=0$ trajectories with $y>0$ must approach the stable focus at late times. Going backwards in time, each such trajectory describes a quasi-cyclic early universe that approaches the heteroclinic cycle more closely in each cycle, as shown in Figs. 3\&4, where (in each of these figures) the lower-hemisphere corresponds to the time-reversed  trajectories with $y<0$. This behaviour was noted in \cite{Billyard:2000cz} for the $D=4$ model with $\lambda = -\mu=2$, except that they chose to consider the case for which the  heteroclinic cycle is approached at {\it late} times;  this happens for $y<0$ but then the Einstein-frame universe is contracting. As we insist on expansion in Einstein frame, we must choose $y>0$, and then the approach to the heteroclinic cycle is at early times, going backwards in time. 
Of particular interest here are the implications for cosmic acceleration. 
The ``acceleration region''  is  the shaded region in Figs. 3\&4.  
\begin{figure}[h!]
  \begin{center}  
      \epsfig{file=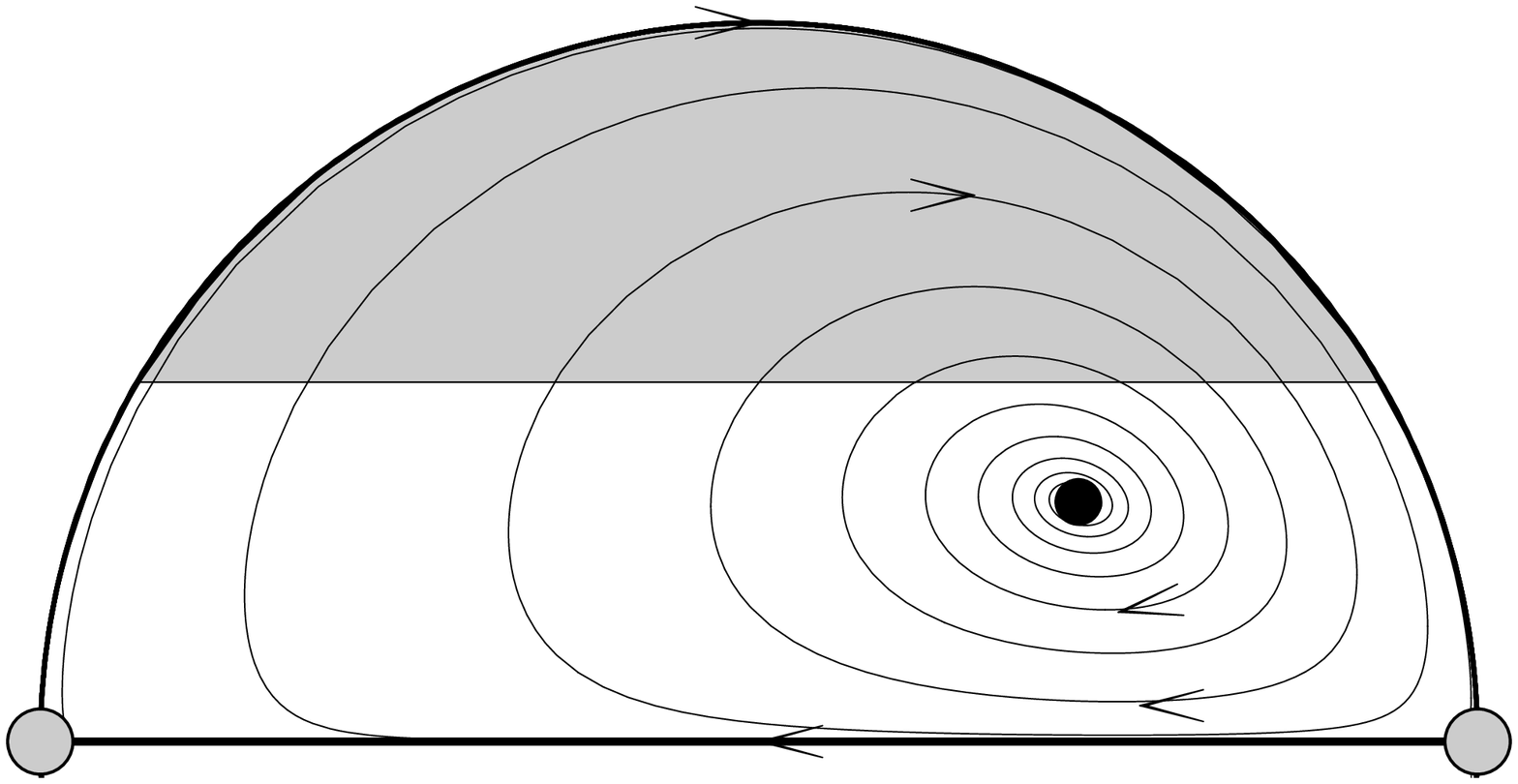, scale=0.3}\\ \vskip1em
	\epsfig{file=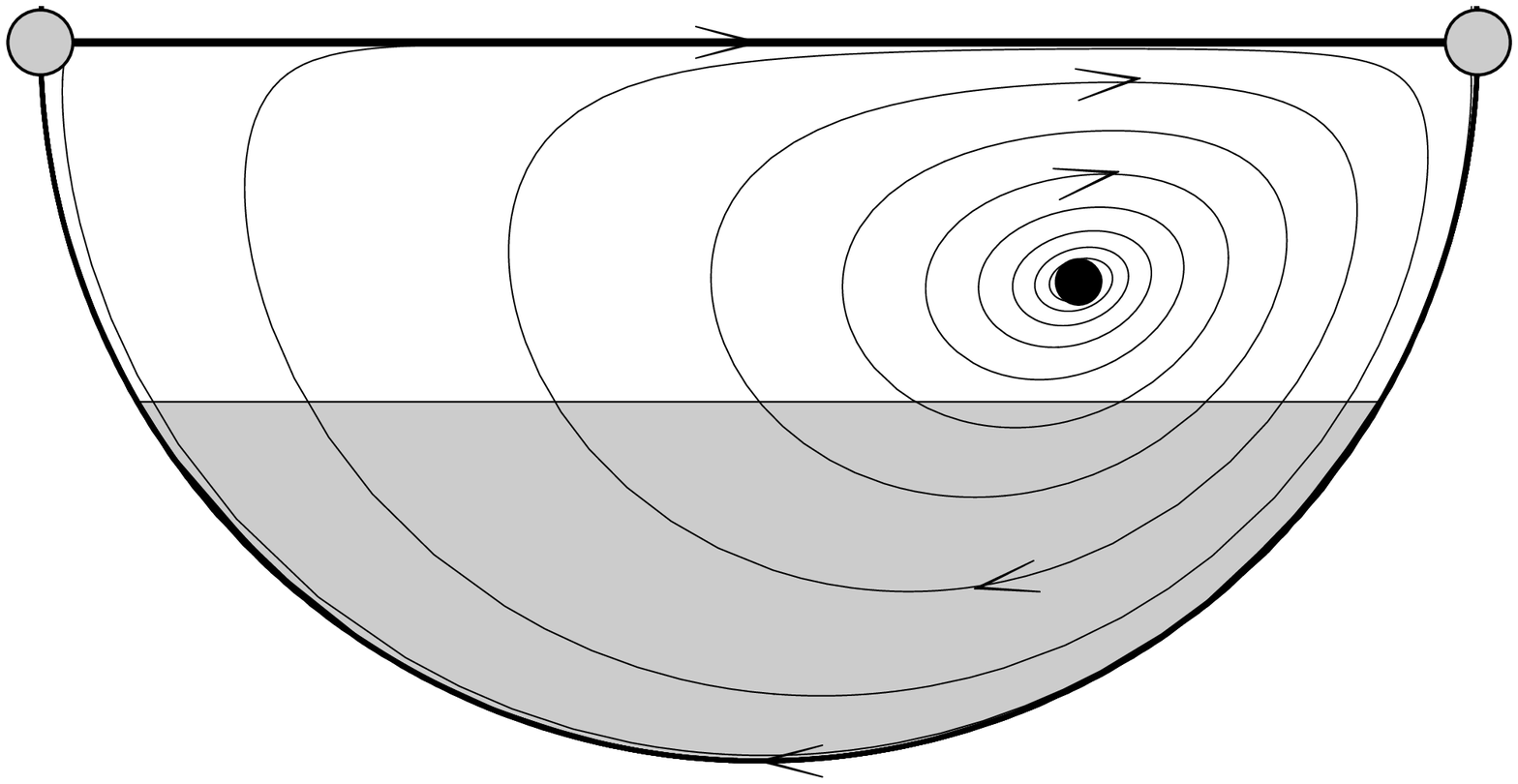,scale=0.3}
      \caption{$\lambda >\lambda_h,\,\mu<\mu_c$, $\mu/\lambda > -2/\lambda_c^2$. Trajectories within the upper (lower) hemicircle represent expanding (contracting) universes.  In each hemicircle, there is a focus {\it outside}  the (shaded) acceleration region. Note the  recurrent acceleration of generic expanding universes.}
      \end{center}

 \end{figure}

The focus
(which is stable for $y>0$ and  unstable for $y<0$) lies \textit{inside} 
the acceleration region if  the ratio $\mu/\lambda$ satisfies 
(\ref{eq:spiral_condition}).  Since the quasi-cyclic trajectories only arise in the range of
parameters for which $\mu_c >0$, any value of $\mu$ that satisfies 
 (\ref{eq:spiral_condition}) will also be such that $\mu<\mu_c$. 
Thus, there exists a range of parameters $(\mu,\lambda)$ such
that there is a stable  focus inside the region of accelerated expansion, and another range that places it  inside the region of decelerated expansion. These two possibilities are illustrated in Figs. 3\&4 respectively, and  the respective parameter ranges are shown in Fig. 5.

\begin{figure}[h!]
      \begin{center}  
      \epsfig{file=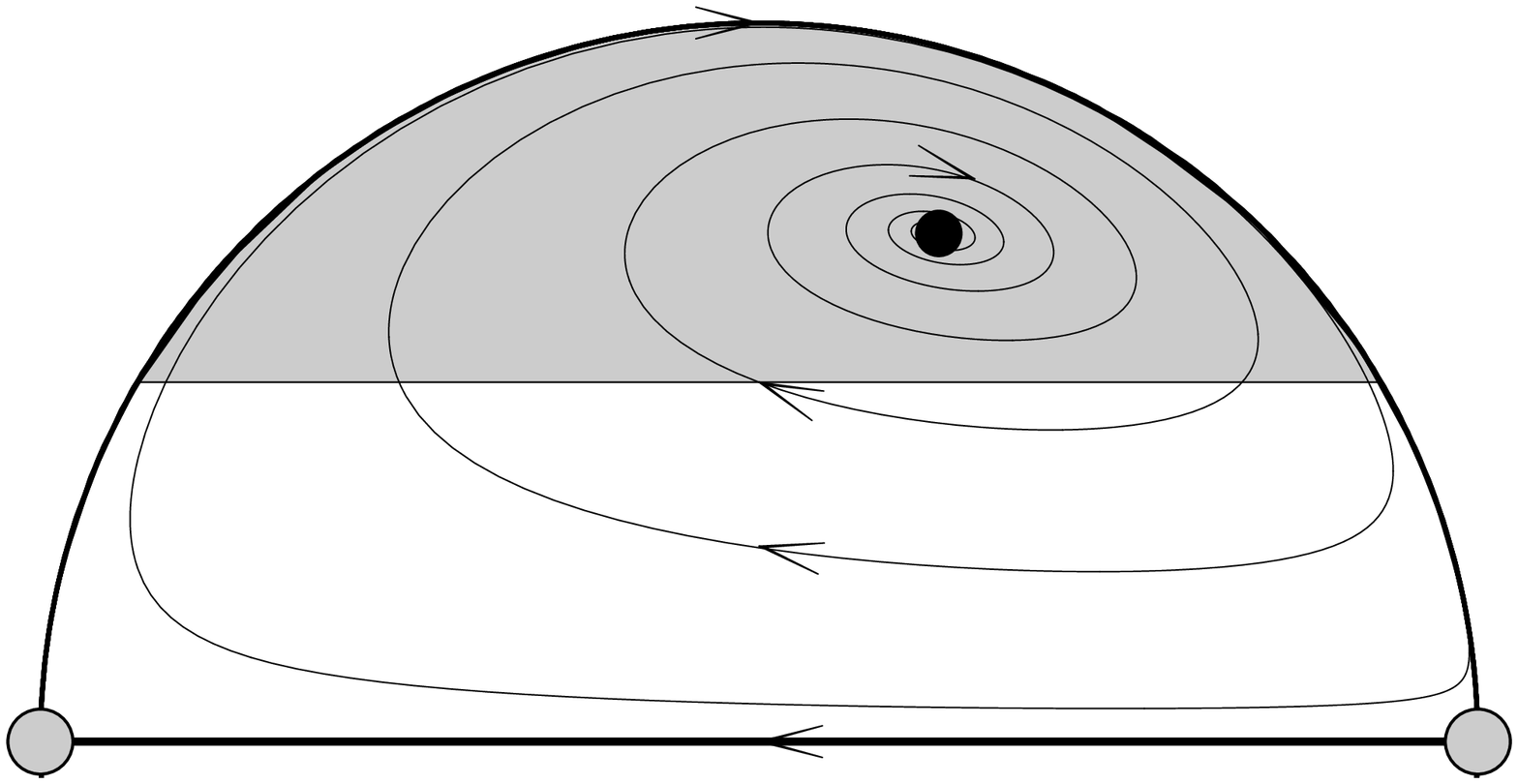, scale=0.3}\\ \vskip1em
	\epsfig{file=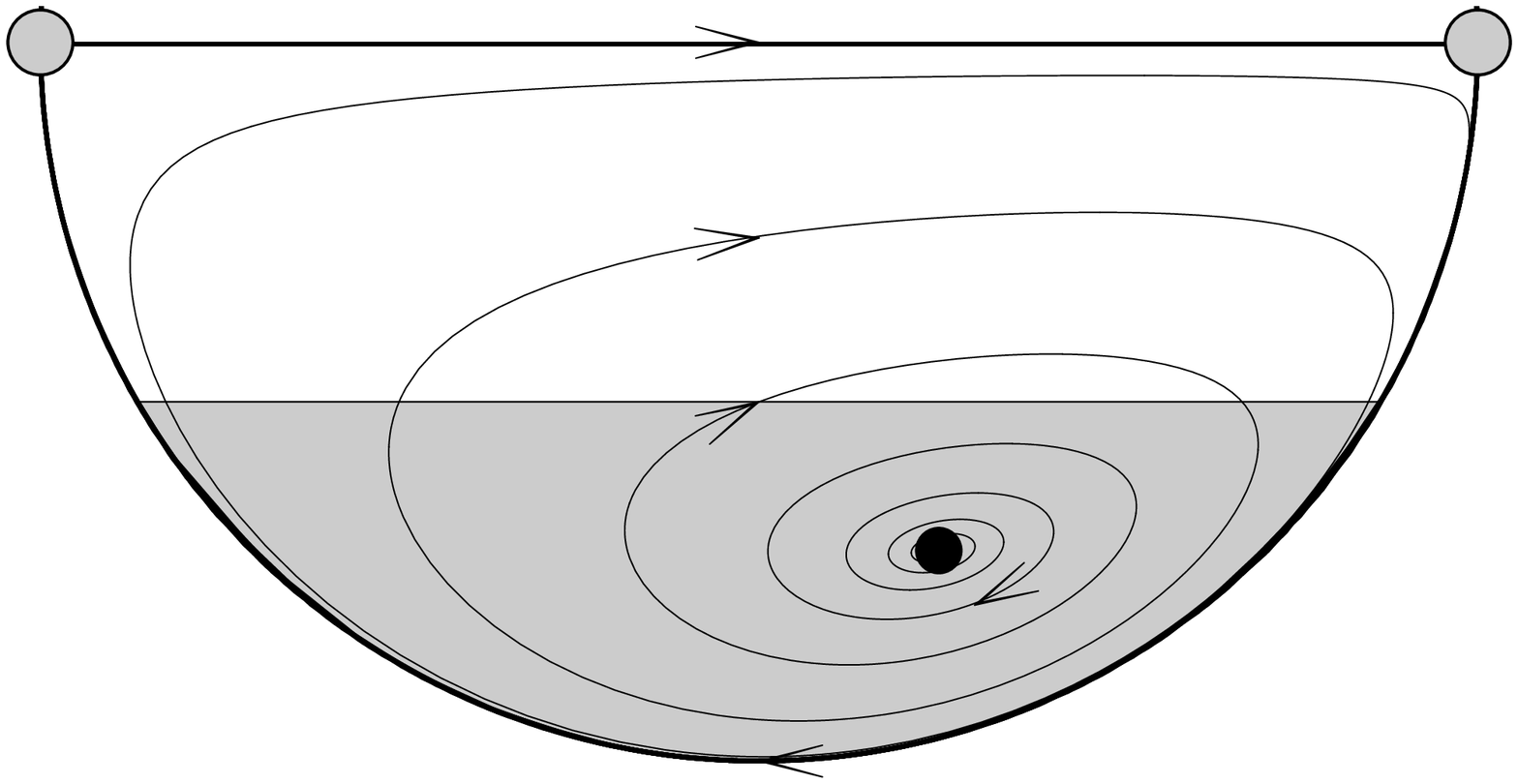,scale=0.3}
      \caption{$\lambda >\lambda_h,\,\mu<\mu_c $, $\mu/\lambda > -2/\lambda_c^2$. For each hemicircle, there is now a focus  {\it  inside} the acceleration region, so that a generic expanding universe undergoes a continuous late-time acceleration, preceeded by early-time recurrent acceleration and deceleration.  \label{fig:cyclicI}}
      \end{center}
  \end{figure}
For generic initial conditions set for very small scale factor (at which either the approximations of general relativity break down or the $k=0$ assumption fails), the universe expands with a series of rapid oscillations between acceleration and deceleration. Potentially, this might have the same effect as an early-time inflationary epoch if the number of cycles of acceleration is sufficiently large, although it must be remembered that the universe under discussion is always flat, by hypothesis, so this ``quasi-inflationary'' epoch could not be used to explain why our Universe is flat. At late times the oscillation between acceleration and deceleration gives way to continuous deceleration, as illustrated in Fig.3, or continuous acceleration, as illustrated in Fig.4.  For the special case that $\mu= -2\lambda/\lambda_c^2$, the stable focus lies on the boundary of the acceleration region and the oscillation between acceleration and deceleration continues forever. 
\begin{figure}[h!]
\begin{center}  
      \epsfig{file=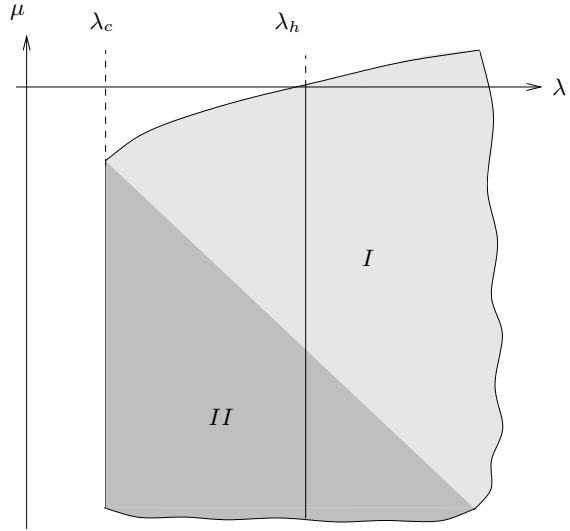, scale=0.7}
\begin{picture}(0,0)(0.01,0.01)
	\put(-72,100){$I$}
	\put(-130,40){$II$}
	\put(0,165){$\lambda$}
	\put(-205,195){$\mu$}
	\put(-105,190){$\lambda_h$}
	\put(-175,190){$\lambda_c$}
	\end{picture}
      \caption{$(\lambda,\mu)$-plane. If the pair of parameters
      $(\mu,\lambda)$ is inside the semi-infinite grey region, there exists a fixed
      point inside the unit circle. If $(\mu,\lambda)$ is in region
      $I$ (light grey), this
      fixed point is outside the acceleration region. If
      $(\mu,\lambda)$ falls into region $II$
      (dark grey) it is located inside the acceleration region.}
      \end{center}
\end{figure}

Finally, we recall that these results depend strongly on the fact 
that we have chosen the Einstein conformal frame. If we define a
conformally rescaled metric 
\be
\tilde g = e^{-r \lambda_c^2\sigma} g
\ee
for arbitrary constant $r$, then we get a Lagrangian density of the
form
\be
{\cal L} = e^{r\sigma}\sqrt{-\det\tilde g}\, \tilde R + \dots\, , 
\ee
and the FLRW scale factor becomes
\be
\tilde S = e^{\beta\varphi - r \lambda_c^2 \sigma}\, . 
\ee
The universe is now expanding in the new conformal frame if $v>
r\lambda_h u$ or, equivalently for $y>0$, $r\lambda_h x<1$.
If $r\lambda_h <1$ then all $y>0$ universes are still eternally
expanding but for $r\lambda_h>1$ there is a late-time oscillation
between expansion and contraction, as observed in
\cite{Billyard:2000cz} for $D=4$ and $r=1$.


\subsection{$\lambda<\lambda_h$}
{}For $\lambda<\lambda_h$ the phase portrait is qualitatively different because of the `finite' fixed point on the boundary of the allowed region. This fixed point (which is the same as the $k=0$ fixed point occurring for $\lambda<\lambda_h$ in the analysis of the $\dot\chi=0$ case of section 3) lies within the acceleration region if $\lambda<\lambda_c$ and outside it if $\lambda>\lambda_c$. We shall consider only the $\lambda>\lambda_c$ case. 

{}For $\lambda<\lambda_h$ we have $\mu_c<0$, and hence, from (\ref{eq:condinf2}), there exists an  `interior'  fixed point when  $\mu<\mu_c$. This fixed point will be a focus if 
\be
\mu>  \mu_- \equiv - \frac{1}{16\lambda} \left[ \lambda_h\sqrt{81\lambda_h^2 - 32\lambda^2} + 
9\lambda_h^2 - 16\lambda^2 \right] \, .
\ee
Note that $\mu_-<\mu_c$ for $\lambda<\lambda_h$, so that there {\it is} a range of values of  $\mu$ for  which this condition is satisfied.  This focus is again stable for $y>0$ and it again lies within the acceleration region if $\mu<-2\lambda/\lambda_c^2$. This condition is satisfied for $\lambda$ sufficiently small, in particular for $\lambda>\lambda_c$ provided that $\lambda-\lambda_c$ is sufficiently small. For larger values of $\lambda<\lambda_h$, the focus must lie in the deceleration region. Thus, given that the late-time behaviour of generic trajectories is governed by an `interior'  
focus, this focus may lie inside or outside the acceleration region, depending on the values of the parameters $(\lambda,\mu)$. The two possibilities are illustrated in Figs 6 \&7.
\begin{figure}[h!]
      \begin{center}  
      \epsfig{file=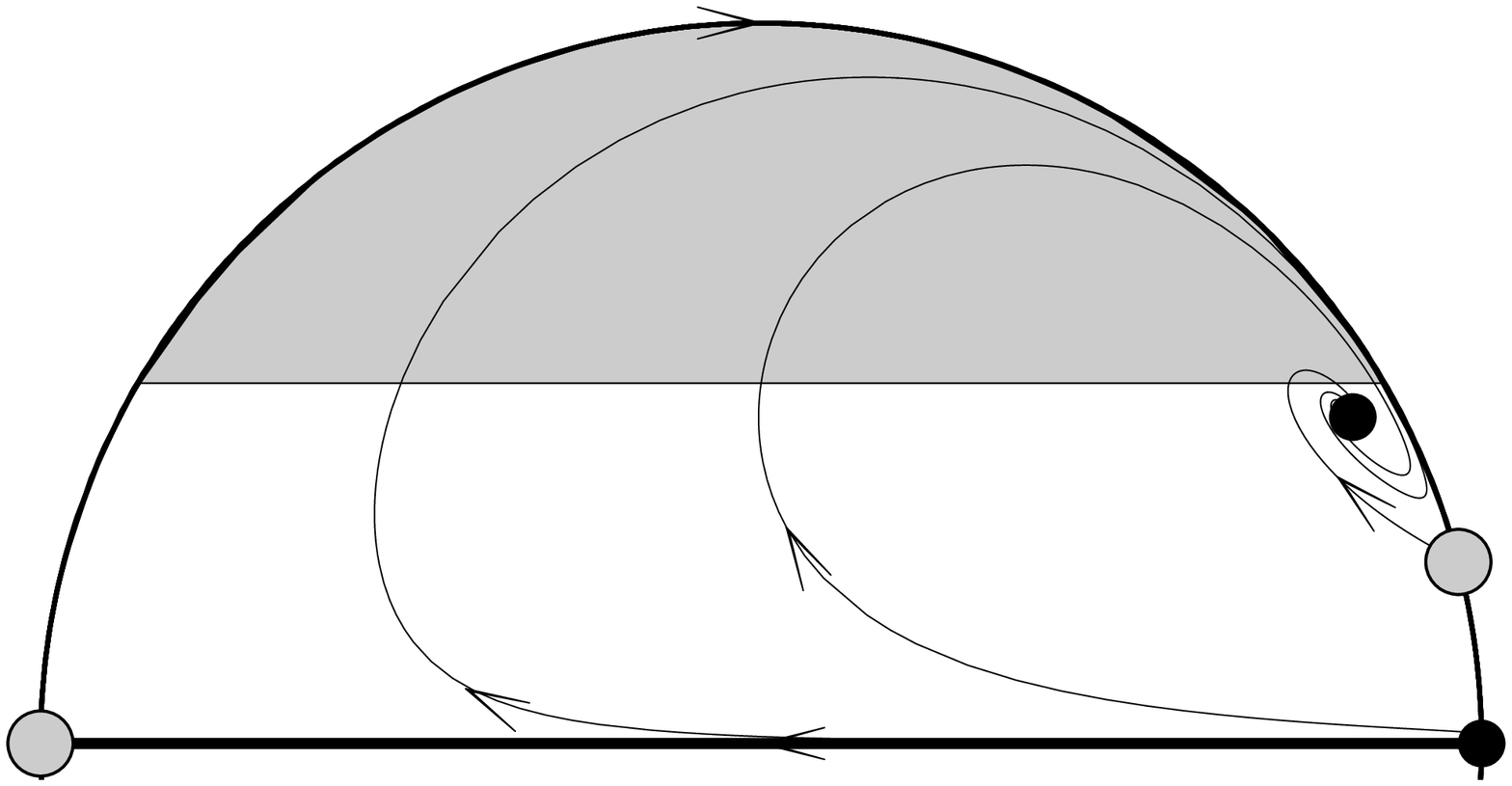, scale=0.3}\\ \vskip1em
	\epsfig{file=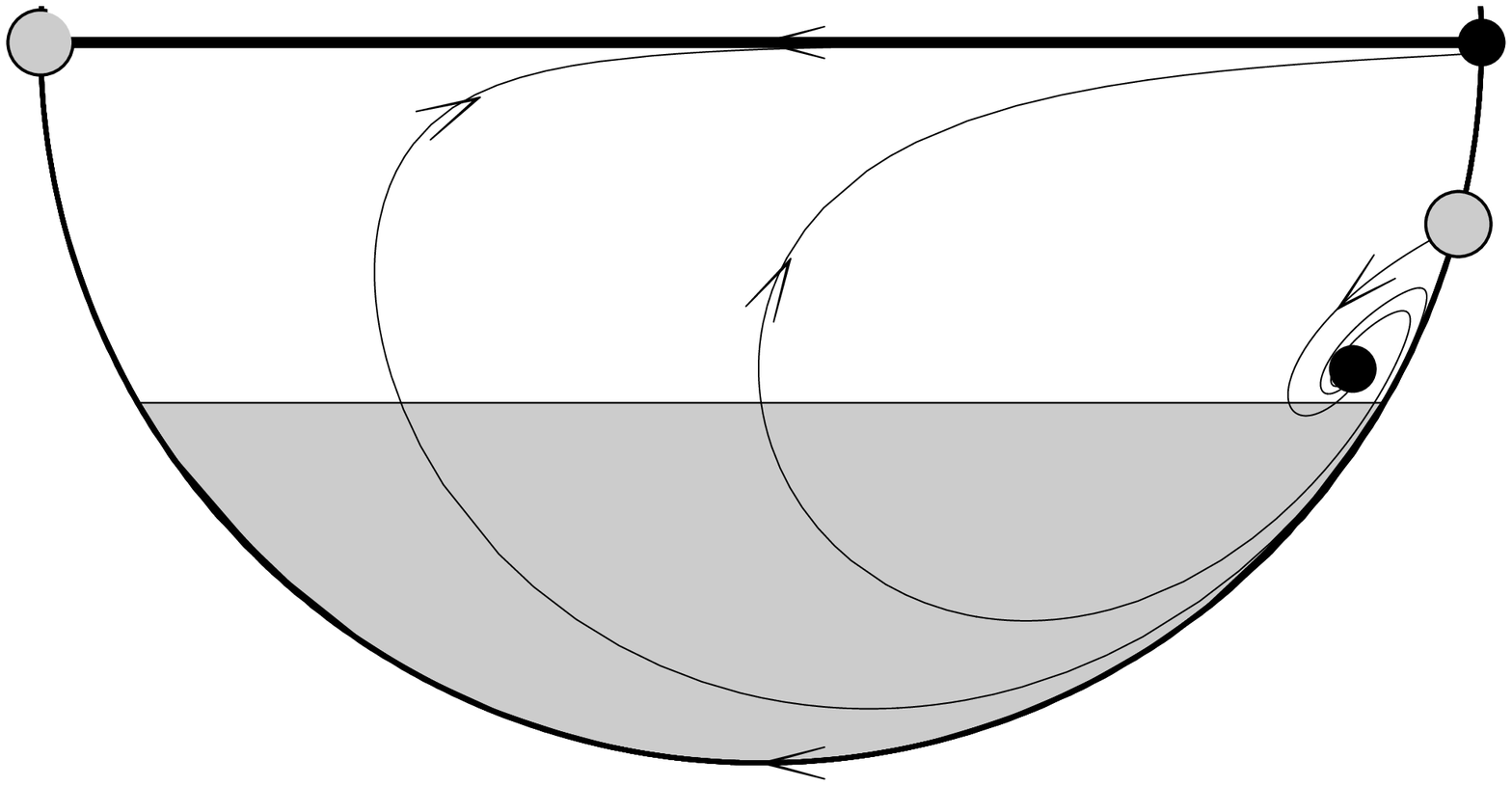,scale=0.3}
      \caption{$\lambda >\lambda_h,\,\mu<\mu_c $, $\mu/\lambda < -2/\lambda_c^2$. 
      Trajectories within the upper (lower) hemicircle represent expanding (contracting) universes.
      Within each hemicircle there is a focus  {\it outside} the acceleration region. }
      \end{center}
  \end{figure}
 Inspection of these phase portraits shows that almost all trajectories begin at the unstable node at $(x,y)=(1,0)$ and end at the `interior' focus (the separatrix trajectory connecting the two fixed points is the unique exception for which $\dot\chi$ is not always zero). Thus, given the existence of the `interior' focus, the late time behaviour for $\lambda_c<\lambda<\lambda_h$ is similar to the late-time behaviour for $\lambda>\lambda_h$. 
\begin{figure}[h!]
    \begin{center}  
      \epsfig{file=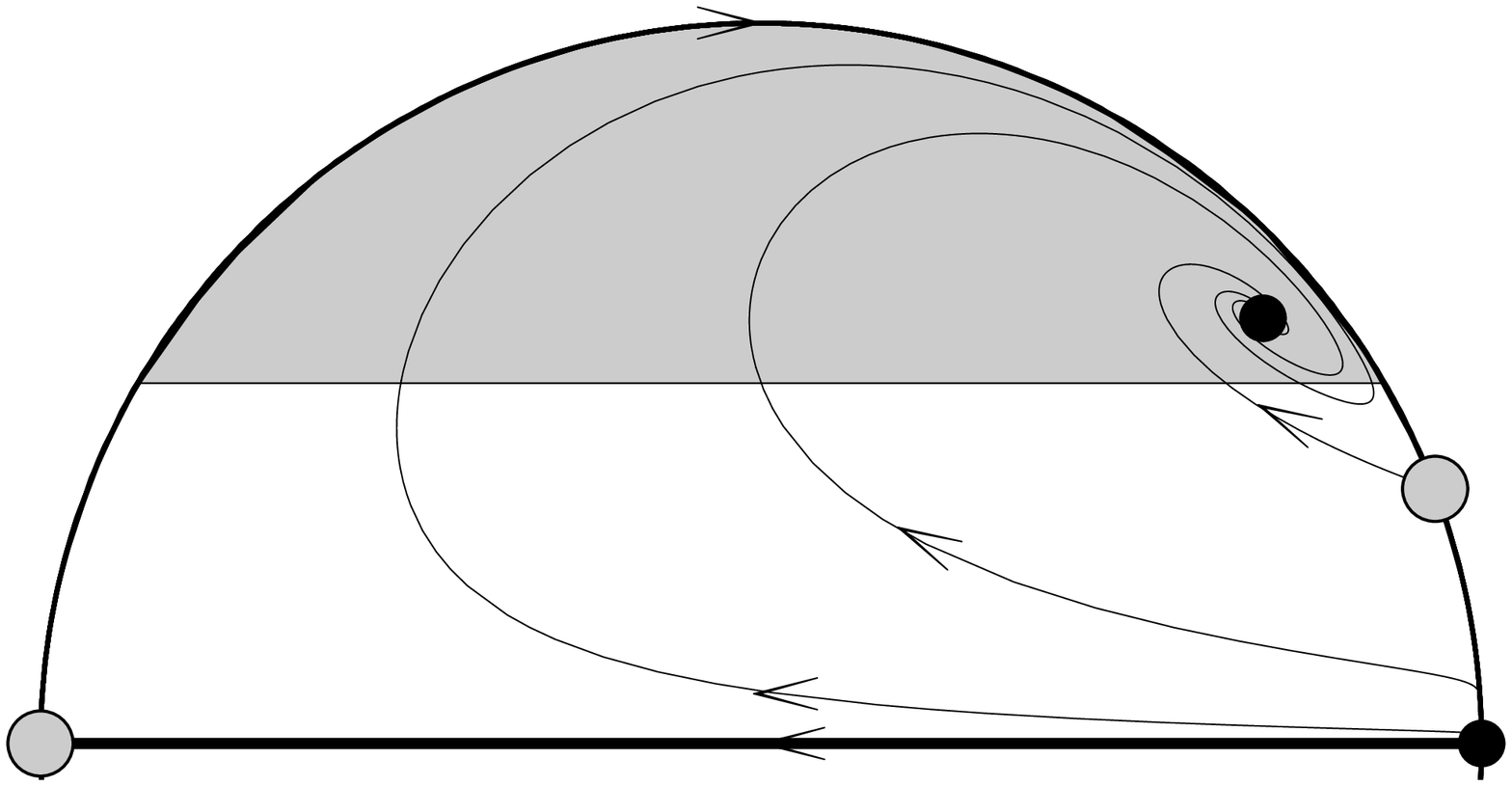, scale=0.3}\\ \vskip1em
	\epsfig{file=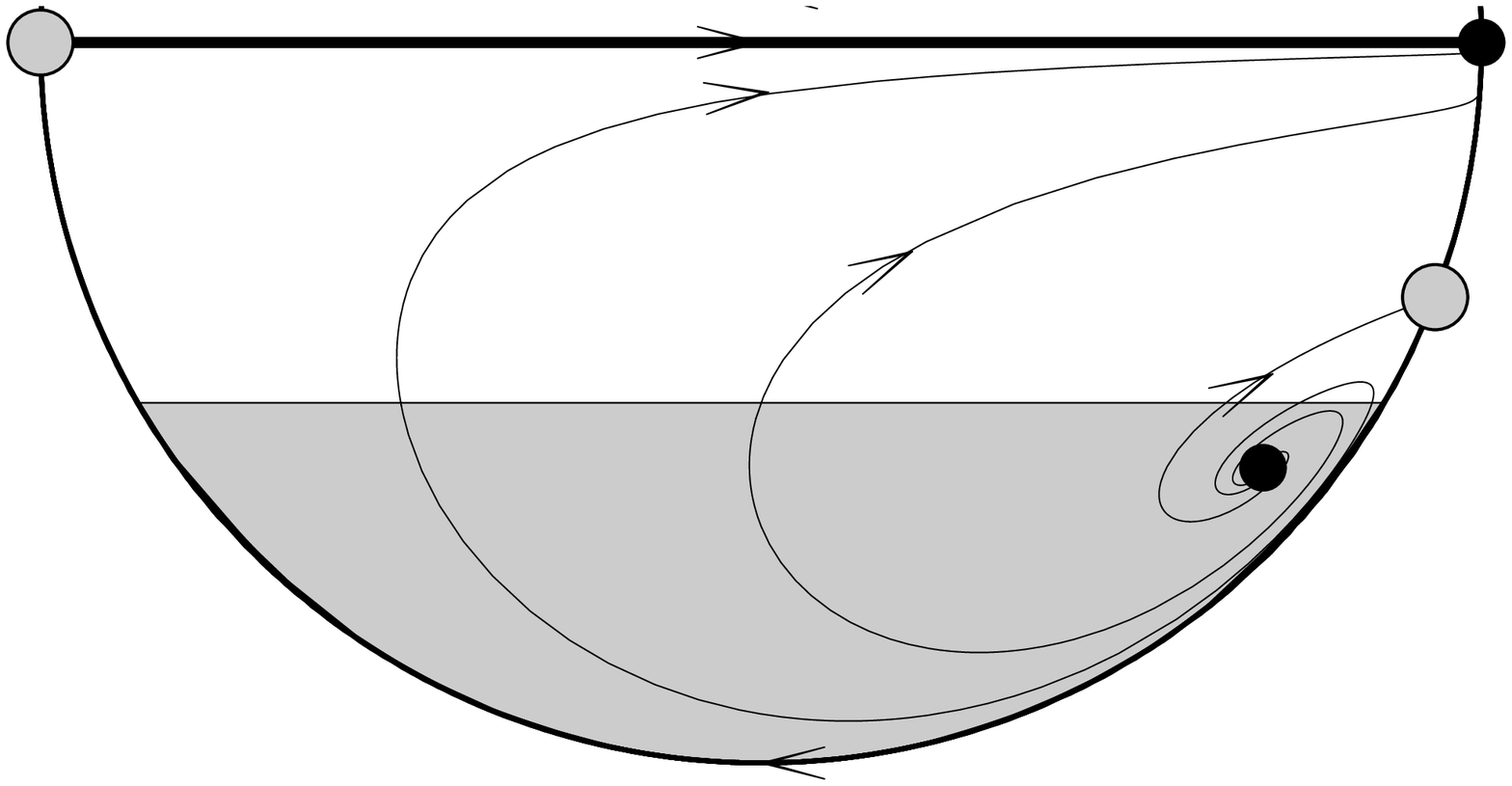,scale=0.3}
      \caption{$\lambda_c<\lambda <\lambda_h,\,\mu<\mu_c$, $\mu/\lambda < -2/\lambda_c^2$. 
      Within each hemicircle there is now a focus {\it inside}  the acceleration region.}
      \end{center}
 \end{figure}

\section{Discussion} \setcounter{equation}{0}

We have presented a study of the homogeneous and isotropic cosmological models that arise in a two-parameter class of models of gravity coupled to dilaton and axion fields in arbitrary spacetime dimension $D$; the two parameters $(\lambda,\mu)$ determine the dilaton coupling constant (and hence the `strength' of an `exponential' potential) and the radius of curvature of the 2-dimensional hyperbolic target space (which we assumed to be finite).  It is well-known that all  possible cosmological solutions of  the zero-axion model can be understood in terms of trajectories of a two-dimensional autonomous dynamical system, but we have explored some new features that arise from consideration of the global nature of the phase space of this system.  It is less well-known that a similar reduction to a two-dimensional autonomous dynamical system is possible for the generic dilaton-axion model provided that the restriction is made to {\it flat}  cosmologies; a particular $D=4$ dilaton-axion model was analysed by this method in \cite{Billyard:2000cz} and we have extended this analysis to the generic dilaton-axion model. In addition, we have explored here the consequences for  {\it Einstein frame} cosmology, which are quite different from those in the `string-inspired' frame of \cite{Billyard:2000cz}. 
In particular, all flat Einstein-frame cosmologies are either eternally expanding or eternally contracting, but may undergo one or more periods of cosmic acceleration.  Our main result is that {\it there is a large range of  the parameters $(\lambda,\mu)$ for which almost all flat expanding universes undergo an epoch of oscillation between acceleration and deceleration}, with either continuous late-time acceleration or continuous late-time deceleration depending on the value of the ratio $\lambda/\mu$ (and eternal oscillation at a critical value of this ratio). 

The early-time behaviour depends strongly on whether $\lambda$ is greater or less than the `hypercritical' value $\lambda_h$. For $\lambda>\lambda_h$, the global phase space is topologically equivalent to the one given in \cite{Billyard:2000cz}, in which almost all trajectories are asymptotic in one direction to a heteroclinic cycle and in the other direction to a focus within this cycle. The focus can be stable or unstable according to the chosen direction of time. In a non-Einstein frame, such as that considered in  \cite{Billyard:2000cz}, for which the universe undergoes quasi-cyclic periods  of expansion and contraction, one may suppose that the focus is unstable and the the heteroclinic cycle is approached at late times. In the Einstein frame, for which the universe is either eternally expanding or eternally contracting, the heteroclinic cycle is approached at late times only by a contracting universe.
If we assume expansion, then the heteroclinic cycle dominates the early-time behaviour, which exhibits oscillation between acceleration and deceleration. This oscillation continues until the focus is approached sufficiently closely that its behaviour, of continuous acceleration or continuous 
deceleration, takes over. This is the origin of the epoch of oscillation between cosmic acceleration and deceleration when $\lambda>\lambda_h$. For $\lambda<\lambda_h$ there is no heteroclinic cycle and the early-time behaviour of expanding universes is quite different, but the medium-time behaviour 
may again exhibit oscillation between cosmic acceleration and deceleration for values of the parameters that put the focus sufficiently close to the borderline between acceleration and deceleration. 

The fact that there are models for which almost all {\it flat} universes undergo cycles of cosmic acceleration and deceleration is of little use for inflationary purposes because one aim of inflation is to explain why the Universe is flat, and for that purpose one needs to consider universes that are not flat. 
In this case, however, the cosmological evolution cannot be reduced to the study of a 2-dimensional phase-plane. It would be of interest to use other methods to learn more about generic non-flat cosmologies (with non-constant axion)  but at present we must assume that there exist some non-flat cosmologies that become approximately flat at late-times, in which case their late-time behaviour  will determine the  `initial' conditions for a flat cosmology. Thus, the early-time behaviour of the flat cosmologies studied in this paper is probably not physically relevant. The late-time behaviour may not be relevant either, depending on whether an almost flat universe becomes more or less flat as time progresses. It is plausible that flatness is a stable feature of universes that exhibit continuous late-time acceleration, and an unstable feature of those that exhibit continuous late-time deceleration. However, the medium-time recurrent acceleration that we have found to be typical of many dilaton-axion models 
could be relevant to the observed cosmic acceleratiion of our Universe. At the very least, it shows that current observations of cosmic acceleration provide no evidence that this acceleration will continue forever.  With many scalar fields, the future is uncertain!
\begin{acknowledgments}
We thank Jonathan Dawes and Aninda Sinha for useful discussions. JS thanks the Gates Cambridge Trust for financial support. PKT thanks the EPSRC for financial support.
\end{acknowledgments}


\end{document}